\documentclass[journal, onecolumn]{IEEEtran}
\usepackage{xcolor}
\ifCLASSINFOpdf
\else
   \usepackage[dvips]{graphicx}
\fi
\usepackage{url}

\usepackage{amsmath,amsfonts}
\usepackage{array}
\usepackage{algorithmic}
\usepackage{xcolor}
\usepackage{algorithm}
\usepackage{array}
\usepackage[caption=false,font=normalsize,labelfont=sf,textfont=sf]{subfig}
\usepackage{textcomp}
\usepackage{stfloats}
\usepackage{url}
\usepackage{verbatim}
\usepackage{graphicx}
\usepackage{cite}

\begin{document}

\title{Fast Computation of the Discrete Fourier Transform Rectangular Index Coefficients}
\author{Saulo Queiroz, João P. Vilela, Benjamin Koon Kei Ng, Chan-Tong Lam, 
 and Edmundo Monteiro, \IEEEmembership{Senior IEEE}
\thanks{Saulo Queiroz (sauloqueiroz@utfpr.edu.br) is with the Academic Department of Informatics of the Federal University of Technology Paran\'a (UTFPR), Ponta Grossa, PR, Brazil, the Centre for Informatics and Systems of the University of Coimbra (CISUC), Portugal.
}
\thanks{Jo\~ao P. Vilela (jvilela@fc.up.pt) is with CRACS/INESCTEC, CISUC and the Department of Computer Science, Faculty of Sciences, University of Porto, Portugal.}
\thanks{Benjamin Koon Kei Ng (bng@mpu.edu.mo) and Chan-Tong Lam (ctlam@mpu.edu.mo) are with the Macao Polytechnic University, Macau, SAR, China.}
\thanks{Edmundo Monteiro (edmundo@dei.uc.pt) is with the Department of Informatics Engineering of the University of Coimbra and CISUC, Portugal.}
\thanks{This work is funded by The Science and Technology Development Fund, Macau SAR. (File no. 0044/2022/A1) 
and Agenda Mobilizadora Sines Nexus (ref. No. 7113), supported by the Recovery and Resilience Plan (PRR) and by the European Funds Next Generation EU.}
}

\markboth{}%
{Saulo Queiroz \MakeLowercase{\textit{et al.}}: Fast Computation of the Discrete Fourier Transform Rectangular Index Coefficients}

\maketitle

\begin{abstract}
In~\cite{sic-magazine-2025}, authors show that the ``square index 
coefficients'' (SIC) of the $N$-point discrete Fourier transform (DFT), 
(i.e., $X_{k\sqrt{N}}$, $k=0,1,\cdots,\sqrt{N}-1$) can be compressed from
$N$ to $\sqrt{N}$ points, thereby speeding up the DFT computation accordingly.
Following up on that, in this article we generalize SICs 
into what we refer to as rectangular index coefficients (RICs) of the DFT,
formalized as  $X_{kL}, k=0,1,\cdots,C-1$, in which the integers $C$ and $L$ 
are generic roots of $N$ such that $N=LC$.
We present an algorithm to compress the $N$-point input signal $\mathbf{x}$ into 
a $C$-point signal $\mathbf{\hat{x}}$ at the expense of $\mathcal{O}(N)$ complex sums
and no complex multiplication. We show that a DFT on $\mathbf{\hat{x}}$
is equivalent to a DFT on the RICs of $\mathbf{x}$. In case where specific 
frequencies of $\mathbf{x}$ are of interest (as turns out to be in harmonic analysis),
one can conveniently adjust the signal parameters (e.g., frequency resolution)
to match RICs to those frequencies and use our algorithm to compute them significantly
faster. If $N$ is a power of two (as usual, due to the  
fast Fourier transform (FFT) algorithm's requirement), then $C$ can be any power of two
in the range $[2, N/2]$ and one can use our algorithm along with the FFT to compute all 
RICs in $\mathcal{O}(C\log C)$ time complexity. 
\end{abstract}

\IEEEpeerreviewmaketitle

\section{Introduction}
\IEEEPARstart{C}{an} the \(N\)-point DFT problem  
be solved faster than the \(\mathcal{O}(N \log_2 N)\) time  
complexity of the FFT algorithm~\cite{CooleyTukey-1965}?  
An answer to this long-standing and still unresolved question  
would have significant implications not only for the  
computational complexity of algorithms~\cite{fftlowerbound-2009}  
but also for some fundamental limits of information theory~\cite{queiroz2024complim,fastenough-2022}.

In many applications~\cite{sfft-survey-2014}, 
signals are ``naturally'' sparse, meaning that several of its DFT
coefficients present (near-)zero amplitude and can be ignored.
Based on this, variants of the FFT algorithm known as sparse FFTs,
claim to perform a reduced number of complex multiplications by avoiding 
the computation of those coefficients. These algorithms, however, modify
the structure of FFT, imposing specific overheads to separate relevant from
negligible coefficients. For this reason, their superior performance against
the traditional FFT is conditioned relatively high values of $N$
and level of sparsity $\kappa$, e.g.,  $\kappa /N=50/2^{22}$~\cite{haitham-soda-2012}.

More recently, in~\cite{sic-magazine-2025}, the authors introduced a form of sparse  
DFT that leverages \(\sqrt{N}\) meaningful coefficients. Unlike other sparse FFT strategies, 
the SIC algorithm losslessly compresses  
\(\sqrt{N}\) DFT coefficients, reducing the input size from \(N\) to \(\sqrt{N}\) points.  
As a result, standard DFT algorithms such as the FFT or the prime-factor  algorithm  (PFA)
can be directly applied to the compressed signal without requiring any modifications -- 
a notable distinction from typical sparse FFT approaches.
By appropriately adjusting signal parameters (e.g., the sampling frequency),  
the SIC DFT algorithm was shown to be effective for relevant applications,
including those related to harmonic analysis.  However, the limitation to only
 \(\sqrt{N}\) coefficients  may be insufficient for other scenarios.

To address this limitation, we propose a generalization of the SIC algorithm  
that allows a flexible number of compressed DFT coefficients, \(C \in [2, N/2]\).  
This approach provides a more precise trade-off between computational complexity  
and (sparse) frequency resolution, thereby extending the algorithm's applicability  
to a broader range of use cases.

The remainder of this work is organized as follows.
In section~\ref{sec:rics}, we formalize the RICs of DFT.
In section~\ref{sec:ricalgorithm}, we present and discuss 
the efficient algorithms to compress RICs and to compute their DFT/IDFTs.
In section~\ref{sec:conclusion}, we present conclusion and
future works.

\section{Rectangular Index Coefficients of DFT}\label{sec:rics}
Let $\mathbf{x}=\{x_0,\cdots,x_{N-1}\}$ denote an $N$-point complex 
signal with DFT $\mathbf{X}=\{X_0,\cdots,X_{N-1}\}$ given by
\begin{eqnarray}
X_{k} &=& \sum_{n=0}^{N-1}x_nW_N^{-kn}, \label{eqn:dft}
\end{eqnarray}
where $W_N$ denote the complex exponential $e^{j2\pi/N}$.
\subsection{Basic Definitions}
For efficient computation using the FFT algorithm, 
$N$ is typically required to be a power of two $2^q$ for some $q=\log_2 N > 0$. 
Note that taking the $q$-th root of $N$ results
in $2$. Based on this, let us define the integer

\begin{eqnarray}
p &\in& [1,q-1]. \label{eqn:p}
\end{eqnarray}
Likewise, let us define the factors of $N$ as follows
\begin{eqnarray}
L &=& \sqrt[q]{N^{q-p}} = \sqrt[q]{(2^q)^{q-p}} = 2^{q-p}, \label{eqn:L}\\
C &=& \sqrt[q]{N^p} = \sqrt[q]{(2^q)^{p}} = 2^p, \label{eqn:C}
\end{eqnarray}
such that
\begin{eqnarray}
N &=& LC. \label{eqn:NCL}
\end{eqnarray}

As we will see next, the formulation of $N$ (\ref{eqn:NCL}) from generic roots of $q$,
given by (\ref{eqn:L}) and (\ref{eqn:C}), will be useful to generalize the index pattern 
of a SIC of the DFT, defined as $X_{k\sqrt{N}}$, $k=0,1,\cdots,\sqrt{N}-1$. 
As shown in~\cite{sic-magazine-2025}, one can exploit the fact that the indexes of SICs 
are multiples of $\sqrt{N}$ to lossely compress them from $N$ to $\sqrt{N}$ 
points. Following up on this, we generalize that idea into what we refer to as RICs, 
defined as $X_{kL}, k=0,1,\cdots,C-1$, thereby enabling such coefficients 
to be compacted from $N$ to $C=\sqrt[q]{N^p}$ points. 

One can freely chose $p$ to set $C$ accordingly.
For $N=2^q$, $C$ can be set to any power of two less than $N$ (if $C=N$ than one gets the 
usual $N$-point DFT). For example, if $N=16=2^4$, then one can have compressed RICs of 
$2^1$, $2^2$, or $2^3$ points. As we will discuss later, the respective values of 
$L$ (namely, $2^3$, $2^2$, and $2^1$) represent the number of
complex additions required by our generalized multiplierless compression algorithm
to compress a single RIC. In any case,  the total complexity to compress all $C$ RICs 
is precisely given by $C\cdot (L-1)=\mathcal{O}(N)$ complex additions and no complex 
multiplication.

In general, the $n$-th 
input sample $x_n$ of the $k$-th DFT coefficient $X_k$ (\ref{eqn:dft}) can 
be arranged into a $L$$\times$$C$ rectangle such that
\begin{eqnarray}
n&=&lC+c \label{eqn:n},
\end{eqnarray}
for $l=0,1,\cdots,L-1$  and $c=0,1,\cdots,C-1$.

\subsection{Compressing RICs}
To show how an $N$-point RIC can be compressed to $C$ points, let us
firstly rewriting (\ref{eqn:dft}) considering (\ref{eqn:n}), this gives
\begin{eqnarray}
X_{k} &=& \left(\sum_{c=0}^{C-1}\sum_{l=0}^{L-1}x_{lC+c}W_N^{-k(lC+c)}\right). \label{eqn:v1} 
\end{eqnarray}
Note that the positions of the summation symbols in (\ref{eqn:v1}) are interchangeable due 
to the commutative property of summations. The order we choose is for the convenience of our 
proof next.
 
Let us consider the case of RICs, i.e., output coefficients $X_{kL}$, 
$k=0,1,\cdots,C-1$. In this case, (\ref{eqn:v1}) rewrites as, 
\begin{eqnarray}
X_{kL} &=& \left(\sum_{c=0}^{C-1}\sum_{l=0}^{L-1}x_{lC+c}W_N^{-kL(lC+c)}\right). \label{eqn:v2} 
\end{eqnarray}
From (\ref{eqn:NCL}), $LC=\sqrt[q]{N^{q-p}}\sqrt[q]{N^{p}}=N$.
Thus, $W_N^{-kLlC}$ results in a root of unit $W_N^{-klN}=1$, and 
\begin{eqnarray}
W_N^{-kLc} &=& W_N^{ k\sqrt[q]{N^{q-p}}c}\nonumber\\
 &=& e^{-j2\pi k{N^{(q-p)/q}}c/N}\nonumber\\
 &=& e^{-j2\pi kc/\sqrt[q]{N^{p}}}\nonumber\\
W_N^{-kLc}&=&W_{C}^{-kc}.
\end{eqnarray}
Based on these observations, (\ref{eqn:v2}) simplifies to 
\begin{eqnarray}
X_{kL} &=& \left(\sum_{c=0}^{C-1}\sum_{l=0}^{L-1}x_{lC+c}W_{C}^{-kc}\right).
\label{eqn:v3} 
\end{eqnarray}

Note that the complex exponential in (\ref{eqn:v3}) is independent of $l$. Based
on this, it results
\begin{eqnarray}
X_{kL} &=& \left(\sum_{c=0}^{C-1}W_{C}^{-kc}\sum_{l=0}^{L-1}x_{lC+c}\right).
\label{eqn:v4} 
\end{eqnarray}

By denoting the inner summation of (\ref{eqn:v4}) as
\begin{eqnarray}
\hat{x}_c &=& \sum_{l=0}^{L-1}x_{l C+c}\label{eqn:xhat},
\end{eqnarray}
Eq.~(\ref{eqn:v4}) rewrites as
\begin{eqnarray}
X_{kL} &=& \sum_{c=0}^{C-1}W_{C}^{{- k c}}{\hat{x}_c},  \label{eqn:compacdft}
\end{eqnarray}
for $k=0,1,\cdots,C-1$.


Note that (\ref{eqn:compacdft}) is a compressed version of the original DFT (\ref{eqn:dft}) 
for RICs. In other words, by performing (\ref{eqn:xhat}) (for $c=0,1,\cdots,C-1$) 
on the $N$-point input signal array $\mathbf{x}=\{x_0,\cdots,x_{N-1}\}$, one gets 
the compressed $C$-point input signal array 
$\mathbf{\hat{x}}=\{\hat{x}_0,\hat{x}_1,\cdots,\hat{x}_{C-1}\}$
at the computational cost of only $\mathcal{O}(N)$ complex sums. 
To compute the output array of coefficients $\mathbf{\hat{X}}=\{\hat{X}_0,\hat{X}_1,\cdots,\hat{X}_{C-1}\}$, 
a $C$-point DFT on $\mathbf{\hat{x}}$ will vary $k$ from $0$ to $C-1$.
Thus, the obtained coefficients match the $C$ SICs of the original input $\mathbf{x}$
following the correspondence,
\begin{eqnarray}
X_{kL}= \hat{X}_{k} \label{eqn:correspondance}
\end{eqnarray}
for $k=0,1,\cdots, C-1$.

 \subsection{Remarks for the Normalized DFT and Inverse DFTs}
As with the original SIC algorithm, the RIC DFT compression
algorithm requires some considerations when applied to normalized or inverse DFTs.  
As is well known, the inverse DFT (IDFT) corresponding to~(\ref{eqn:dft}) is given by  
\begin{eqnarray}
x_{n} &=& \frac{1}{N}\sum_{k=0}^{N-1}X_kW_N^{kn}. \label{eqn:idftn}
\end{eqnarray}
for $n=0,1,\cdots, N-1$.
Following the same reasoning used for the DFT RICs,
the corresponding IDFT computation of the RICs is given by
\begin{eqnarray}
\hat{x}_{n} &=& \frac{1}{C}\sum_{c=0}^{C-1}{\hat{X}_c}W_{C}^{{ n c}},  \label{eqn:compactidft}
\end{eqnarray}
for $n=0,1,\cdots,C-1$ and 
\begin{eqnarray}
\hat{X}_c &=& \sum_{l=0}^{L-1}X_{l C+c}\label{eqn:Xhat},
\end{eqnarray}
for $c=0,1,\cdots, C-1$. In other words, the  input and output of
the $C$-point compressed IDFT are the signals 
$\mathbf{\hat{X}}=\{\hat{X}_0,\cdots, \hat{X}_{C-1}\}$ and
 $\mathbf{\hat{x}}=\{\hat{x}_0,\cdots, \hat{x}_{C-1}\}$,
respectively. However, the correspondance between 
an $N$-point RIC $x_{nL}$ ($n=0,\cdots,C-1$) and its compressed counterpart $\hat{x}_n$
is not given straightforwardly as in the DFT case (\ref{eqn:correspondance}), 
 because they have the distinct normalization factors
$N^{-1}$ and $C^{-1}$, respectively. Considering this point and recalling that $N=LC$, 
the following correspondance holds,
\begin{eqnarray}
x_{nL}= \frac{1}{L}\hat{x}_{n} \label{eqn:correspondancen}
\end{eqnarray}
for $n=0,\cdots,C-1$.
In other cases, one may also normalize (\ref{eqn:dft}) to $(\sqrt{N})^{-1}$
in order to preserve the signal's energy after the transform. In this case,
the inverse DFT is given by
\begin{eqnarray}
x_{n} &=& \frac{1}{\sqrt{N}}\sum_{k=0}^{N-1}X_kW_N^{kn}, \label{eqn:idftsqrtn}
\end{eqnarray}
for $n=0,\cdots,N-1$, and the equivalent compressed $C$-point DFT is
\begin{eqnarray}
\hat{x}_{n} &=& \frac{1}{\sqrt{C}}\sum_{c=0}^{C-1}{\hat{X}_c}W_{C}^{{ k c}}. \label{eqn:compactidftsqrt}
\end{eqnarray}
for $n=0,\cdots,C-1$.
As before,  the correspondance between $x_{nL}$ and $\hat{x}_n$ must account for the
normalization factor, which is given by
\begin{eqnarray}
x_{nL}= \frac{1}{\sqrt{L}}\hat{x}_{n} \label{eqn:correspondancensqrt}
\end{eqnarray}
for $n=0,1,\cdots, C-1$, since $(1/\sqrt{L})(1/\sqrt{C})=1/\sqrt{N}$.

\begin{algorithm}[t]
\caption{RIC DFT (IDFT) Algorithm.\label{alg:ricdft}}
\begin{algorithmic}[1]

\STATE \textbf{RIC\_DFT($N$, $\mathbf{x}$, $sign$, $K$)} 
\STATE \textbf{Input:} $N=LC$-point signal array $\mathbf{x}$, \textbf{$sign$}: $-1$ for DFT, $+1$ for IDFT,
$K$ normalization factor, i.e., $1$ or $1/\sqrt{N}$ for DFT,
 $1/N$ or $1/\sqrt{N}$ for IDFT.

\STATE \textbf{Output:} \text{$C$-point transform $X_{kL}$, \text{for $k=0,1,\cdots, C-1$}};

\STATE \label{ln:compression} $\mathbf{\hat{x}}\gets \textbf{RIC\_Compression($N$, $\mathbf{x}$)}$; \COMMENT{Compressing from $N$-point to $C$-point signal with Algorithm \ref{alg:compressric}}

\IF{\textbf{$sign$} is $-1$} \label{ln:sign}
    \STATE $\mathbf{\hat{X}} \gets \mathcal{F}(\mathbf{\hat{x}})$;\label{ln:dft}
    \STATE $K \gets 1$; \COMMENT{Here, $K\mathbf{\hat{X}}$ gives $\mathbf{\hat{X}}$}
\ELSE
    \STATE $\mathbf{\hat{X}} \gets \mathcal{F}^{-1}(\mathbf{\hat{x}})$; \label{ln:idft}
    \IF{\textbf{$K$} is $\frac{1}{N}$}
        \STATE $K \gets \frac{1}{L}$; \COMMENT{Here, $K\mathbf{\hat{X}}$ gives $\frac{1}{N}\mathbf{\hat{X}}$}\label{ln:idft1l}
    \ELSE
        \STATE $K \gets \frac{1}{\sqrt{L}}$; \COMMENT{Here, $K\mathbf{\hat{X}}$ gives $\frac{1}{\sqrt{N}}\mathbf{\hat{X}}$}\label{ln:idft1sqrtl}
    \ENDIF
\ENDIF
\STATE $X_{kL}\gets K\hat{X}_k$, \text{for $k=0,1,\cdots, C-1$ and $\hat{X}_k\in \mathbf{\hat{X}}$}\label{ln:fixnorm}
\RETURN $X_{kL}$, \text{for $k=0,1,\cdots, C-1$};
\end{algorithmic}
\end{algorithm}

\section{Algorithms}\label{sec:ricalgorithm}
Algorithm~\ref{alg:ricdft} computes the DFT (or IDFT)
 $\mathbf{\hat{X}}$ for the RICs of an $N$-point input 
signal $\mathbf{x}$. As previously explained, if the integer $N$ 
is a power of two, then $C$ (hence $L$) can be any power of two
in the range $[2, N/2]$. The input parameter \texttt{sign} denotes 
the sign of the complex exponentials in the DFT.  
Accordingly, a value of $-1$ indicates that a DFT is to be 
performed, while $+1$ indicates an IDFT. Finally, the input parameter 
$K$ denotes the normalization implicitely applied by the DFT and IDFT
procedures $\mathcal{F}$ (ln. \ref{ln:dft}) and 
$\mathcal{F}^{-1}$ (ln. \ref{ln:idft}),
respectively.

\subsection{Description}
Algorithm~\ref{alg:ricdft} begins by compressing the $N$-point
input signal $\mathbf{x}$ to the equivalent $C$-point signal 
$\mathbf{\hat{x}}$ (ln.~\ref{ln:compression}). This is performed 
using the ``RIC\_Compression'' Algorithm \ref{alg:compressric},
which directly results from the equations derived in Section~\ref{sec:rics}.

Line~\ref{ln:sign} determines whether a DFT or an IDFT will be performed,
as previously explained. Commonly, IDFT implementations normalize the output coefficients  
either by the reciprocal of the signal length or by the reciprocal of its square root.  
Given the \(C\)-point signal provided as input to the IDFT operation  
\(\mathcal{F}^{-1}(\mathbf{\hat{x}})\) in ln.~\ref{ln:idft}, these respective  
normalizations would result in \(1/C\) and \(1/\sqrt{C}\), rather than the  
expected \(1/N\) and \(1/\sqrt{N}\). 
To ensure that the \(N\)-point normalizations hold for the final computed RIC coefficients,  
the RIC IDFT algorithm applies an additional scaling factor \(K\) (ln.~\ref{ln:fixnorm}).  
This factor is set to \(K = 1/L\) (ln.~\ref{ln:idft1l}), since \((1/L)(1/C) = 1/N\),  
or to \(K = 1/\sqrt{L}\) (ln.~\ref{ln:idft1sqrtl}), since \((1/\sqrt{L})(1/\sqrt{C}) = 1/\sqrt{N}\),  
depending on the chosen normalization scheme implicit to the IDFT implementation \(\mathcal{F}^{-1}\) 
in ln.~\ref{ln:idft}.
Analogously, the DFT coefficients are correctly handled when a normalization factor of  
\(1/\sqrt{N}\) is implicitly applied by the DFT implementation \(\mathcal{F}\) in ln.~\ref{ln:dft}.

\begin{algorithm}[t]
\caption{RIC Compression Algorithm.\label{alg:compressric}}
\begin{algorithmic}[1]
\STATE \textbf{RIC\_Compression($N$, $\mathbf{x}$)} 
\STATE \textbf{Input:} $N=LC$-point signal array $\mathbf{x}$
\STATE \textbf{Output:} $C$-point compressed array $\mathbf{\hat{x}}$
\STATE \text{allocate the vector} $\hat{\mathbf{x}}[0,\cdots,C-1]\gets 0$;
\FOR{$c = 0$ \TO $C - 1$}
    \STATE $\hat{\mathbf{x}}[c] \gets 0$;
    \FOR{$l = 0$ \TO $L - 1$}
        \STATE $\hat{\mathbf{x}}[c] \gets \hat{\mathbf{x}}[c] + \mathbf{x}[c + lC]$;\label{ln:ricsums}
    \ENDFOR
\ENDFOR
\RETURN $\hat{\mathbf x}$;
\end{algorithmic}
\end{algorithm}

\subsection{Computational Complexity}
The asymptotic complexity of Algorithm~\ref{alg:ricdft} is primarily determined  
by the computational costs of the $\textbf{RIC\_Compression}$ (ln.~\ref{ln:compression}) 
and DFT/IDFT procedures (lns.~\ref{ln:dft} and \ref{ln:idft}, respectively).

The complexity of the RIC compression procedure (Algorithm~\ref{alg:compressric})  
results from two nested loops with \(C\) and \(L\) iterations, respectively.  
Since \(N = CL\), the total number of iterations is \(\mathcal{O}(N)\).  
It is worth noting that this cost accounts only for complex additions (ln.~\ref{ln:ricsums}),  
as no complex multiplications are performed.  
The complexities of the DFT/IDFT procedures, in turn, depend on the specific algorithms used.  
In general, if the DFT/IDFT has a complexity of \(T(N)\) for an \(N\)-point signal,  
then the transforms in lns.~\ref{ln:dft} and~\ref{ln:idft} will each have a complexity of \(T(C)\).
Assuming \(C\) satisfies the requirements of algorithms such as the FFT or PFA,  
these algorithms can be employed without modification, yielding a complexity of \(\mathcal{O}(C \log_2 C)\).  
Recall that, if \(N\) is a power of two, then \(C\) can be chosen as any power of two  
in the range \([2, N/2]\), as previously discussed.
Therefore, the total cost to compute all $C$ $N$-point RICs is $\mathcal{O}(N)$ 
complex additions plus $\mathcal{O}(C\log_2 C)$ complex multiplications.

It is also worth noting that the SIC algorithm~\cite{sic-magazine-2025}  
is a particular case of the proposed RIC algorithm when \(L = C\).  
Moreover, if frequency resolution is not critical for the application,  
the SIC configuration is preferable, as it yields the lowest number of  
complex multiplications—namely, \(\mathcal{O}(\sqrt{N} \log_2 \sqrt{N})\).

\subsection{DFT Example}\label{sec:dftexample}
Consider the following example of a $N=8$-point signal,
\begin{eqnarray}
\mathbf{x} &=&\{1+1j, 2+2j, 3+3j, -4-4j,\nonumber\\
              && -5-5j, -6+6j,  7-7j, 8+8j\}.\label{eqn:xexample}
\end{eqnarray}
The possible $L\times C$ setups are $2^1\times 2^2$ ($C=4$) and $2^2\times 2^1$ ($C=2$).
For this example, we choose to set $C=4$, hence $L=2$. 
The first step consisting in computing the multiplierless summation 
(\ref{eqn:xhat}), which represents the lossless compression of our
algorithm. It results from adding the samples of 
$\mathbf{x}$ at every $C=4$ step to get the $C=4$-point
vector $\mathbf{\hat{x}}$. This yields,
\begin{eqnarray}
\hat{x}_0&=&\sum_{l=0}^{2-1}x_{4l+0}=(1+1j)-(5-5j)=-4-4j,\nonumber\\
\hat{x}_1&=&\sum_{l=0}^{2-1}x_{4l+1}=(2+2j)-(6+6j)=-4+8j,\nonumber\\
\hat{x}_2&=&\sum_{l=0}^{2-1}x_{4l+2}=(3+3j)+(7-7j)=10-4j,\nonumber\\
\hat{x}_3&=&\sum_{l=0}^{2-1}x_{4l+3}=(-4-4j)+(8+8j)=4+4j,\nonumber
\end{eqnarray}
therefore,
\begin{eqnarray}
\mathbf{\hat{x}} =\{-4-4j, -4+8j, 10-4j, 4+4j\}\label{eqn:xhatexample}
\end{eqnarray}

Note that the compressed signal $\mathbf{\hat{x}}$ was obtained from $\mathbf{x}$
by performing only $L-1$  complex additions.
A DFT on $\mathbf{\hat{x}}$ will produce the output signal 
vector $\mathbf{\hat{X}}=\{\hat{X}_0,\hat{X}_1,\hat{X}_2, \hat{X}_3\}$.
Recall that $X_{kC}=\hat{X}_k$ for $k=0,1,\cdots, C-1$ (\ref{eqn:correspondance}).
Therefore, by performing a DFT on $\mathbf{\hat{x}}$, one obtains
the RICs from the compressed 4-point signal for $L=2$ as follows:
\begin{eqnarray}
X_{0L} &=& \hat{X}_0  = \sum_{c=0}^{4-1}W_{4}^{{- 0\cdot c}}{\hat{x}_c}\approx 6 + 4j,  \nonumber\\
X_{1L} &=& \hat{X}_1  = \sum_{c=0}^{4-1}W_{4}^{{- 1\cdot c}}{\hat{x}_c}\approx -10 + 8j,  \nonumber\\
X_{2L} &=& \hat{X}_2  = \sum_{c=0}^{4-1}W_{4}^{{- 2\cdot c}}{\hat{x}_c}\approx 6 -20j,\nonumber\\
X_{3L} &=& \hat{X}_3  = \sum_{c=0}^{4-1}W_{4}^{{- 3\cdot c}}{\hat{x}_c}\approx -18-8j.\nonumber
\end{eqnarray}
resulting in 
\begin{eqnarray}
\mathbf{\hat{X}} =\{6 + 4j, 10 + 8j, 6 -20j, 18-8j\}.\label{eqn:Xhatexample}
\end{eqnarray}

\subsection{IDFT Example}
Consider the case study of section~\ref{sec:dftexample} 
for the computation of the IDFT RICs  instead of the DFT.
In this case, the same $C=4$-point compressed signal of the DFT example
(\ref{eqn:xhatexample}) results, 
\begin{eqnarray}
\mathbf{\hat{X}} =\{-4-4j, -4+8j, 10-4j, 4+4j\}.\label{eqn:Xhatidftexample}
\end{eqnarray}

By performing the 
$C$-point IDFT (\ref{eqn:compactidft}) on $\mathbf{\hat{X}}$ one gets,
 \begin{eqnarray}
\mathbf{\hat{x}} &=&\{1.5 +j, -4.5 -2j, 1.5 -5j, -2.5 + 2j\}.\label{eqn:xhatidftexample}
\end{eqnarray}

Assuming the chosen normalization must correspond to the reciprocal of the signal
length, the IDFT coefficients must be multiplied by $1/N=1/8$ rather than $1/C=1/4$. To
address this, one multiplies the compressed computed signal to $L^{-1}$ according to (\ref{eqn:correspondancen}).
Since $L=2$ in this case, the final $8$-point RICs result
 \begin{eqnarray}
x_{0L} &=&  \frac{1}{2}\hat{x}_{0} \approx 0.75 +0.5j,  \nonumber\\
x_{1L} &=& \frac{1}{2}\hat{x}_{1} \approx -2.25 -j,  \nonumber\\
x_{2L} &=& \frac{1}{2}\hat{x}_{2}\approx 0.75 -2.5j,\nonumber\\
x_{3L} &=& \frac{1}{2}\hat{x}_{3}\approx -1.25 + j.\nonumber
\end{eqnarray}

\section{Conclusion}\label{sec:conclusion}
In this work, we defined the RICs of an \(N = 2^q\)-point DFT and 
demonstrated that they can be losslessly compressed from \(N\) to \(C\) 
points, where \(C\) is any power of two less than \(N\). A natural direction 
for future work is to study the roundoff errors associated with the \(C\)-point 
RICs and compare them to those of the original \(N\)-point coefficients. Another 
important avenue is to investigate whether substituting \(N\)-point computations 
with \(C\)-point RICs impacts the spectral leakage characteristics of the DFT. 
Finally, a more challenging problem lies in identifying patterns that would enable 
the lossless compression of odd-indexed DFT coefficients.

\bibliographystyle{IEEEtran}
\bibliography{IEEEabrv,refs}

\begin{thebibliography}{1}
\providecommand{\url}[1]{#1}
\csname url@samestyle\endcsname
\providecommand{\newblock}{\relax}
\providecommand{\bibinfo}[2]{#2}
\providecommand{\BIBentrySTDinterwordspacing}{\spaceskip=0pt\relax}
\providecommand{\BIBentryALTinterwordstretchfactor}{4}
\providecommand{\BIBentryALTinterwordspacing}{\spaceskip=\fontdimen2\font plus
\BIBentryALTinterwordstretchfactor\fontdimen3\font minus
  \fontdimen4\font\relax}
\providecommand{\BIBforeignlanguage}[2]{{%
\expandafter\ifx\csname l@#1\endcsname\relax
\typeout{** WARNING: IEEEtran.bst: No hyphenation pattern has been}%
\typeout{** loaded for the language `#1'. Using the pattern for}%
\typeout{** the default language instead.}%
\else
\language=\csname l@#1\endcsname
\fi
#2}}
\providecommand{\BIBdecl}{\relax}
\BIBdecl

\bibitem{sic-magazine-2025}
S.~Queiroz, J.~ao~P.~Vilela, and E.~Monteiro, ``Fast computation of the
  discrete fourier transform square index coefficients,'' \emph{scheduled for
  publication in IEEE Signal Processing Magazine (Tips \& Tricks)}, 2025.

\bibitem{CooleyTukey-1965}
J.~Cooley and J.~Tukey, ``An algorithm for the machine calculation of complex
  fourier series,'' \emph{Mathematics of Computation}, vol.~19, no.~90, pp.
  297--301, 1965.

\bibitem{fftlowerbound-2009}
S.~V. Lokam, \emph{Complexity Lower Bounds Using Linear Algebra}.\hskip 1em
  plus 0.5em minus 0.4em\relax Hanover, MA, USA: Now Publishers Inc., 2009.

\bibitem{queiroz2024complim}
S.~Queiroz, J.~P. Vilela, and E.~Monteiro, ``Computation-limited signals: A
  channel capacity regime constrained by computational complexity,'' \emph{IEEE
  Communications Letters}, vol.~28, no.~8, pp. 1909--1913, 2024.

\bibitem{fastenough-2022}
------, ``{Is FFT Fast Enough for Beyond 5G Communications? A
  Throughput-Complexity Analysis for OFDM Signals},'' \emph{IEEE Access},
  vol.~10, pp. 104\,436--104\,448, 2022.

\bibitem{sfft-survey-2014}
A.~C. Gilbert, P.~Indyk, M.~Iwen, and L.~Schmidt, ``Recent developments in the
  sparse fourier transform: A compressed fourier transform for big data,''
  \emph{IEEE Signal Process. Magazine}, vol.~31, no.~5, pp. 91--100, Sept 2014.

\bibitem{haitham-soda-2012}
H.~Hassanieh, P.~Indyk, D.~Katabi, and E.~Price, ``Simple and practical
  algorithm for sparse fourier transform,'' in \emph{Proceedings of the
  Twenty-Third Annual ACM-SIAM Symposium on Discrete Algorithms}, ser. SODA
  '12.\hskip 1em plus 0.5em minus 0.4em\relax USA: Society for Industrial and
  Applied Mathematics, 2012, pp. 1183--1194.

\end{thebibliography}

\end{document}